\documentclass{jcgt}

\setciteauthor{Malyshau}
\setcitetitle{Six Ways to Draw Vangers with WebGPU: Real-Time Rendering of Editable Multi-Layer Height Fields}
\setheadtitle{Six Ways to Draw Vangers with WebGPU}
\submitted{2026-08-17}

\usepackage{longtable}
\usepackage{booktabs}
\usepackage{array}
\usepackage{tabularx}
\newcolumntype{Y}{>{\raggedright\arraybackslash}X}
\providecommand{\tightlist}{\setlength{\itemsep}{0pt}\setlength{\parskip}{0pt}}
\providecommand{\pandocbounded}[1]{#1}
\providecommand{\passthrough}[1]{#1}

\begin{document}

\title{Six Ways to Draw Vangers with WebGPU:\\Real-Time Rendering of Editable Multi-Layer Height Fields}

\author{Dzmitry Malyshau~\href{https://orcid.org/0009-0005-6410-4276}{\includegraphics[width=8pt]{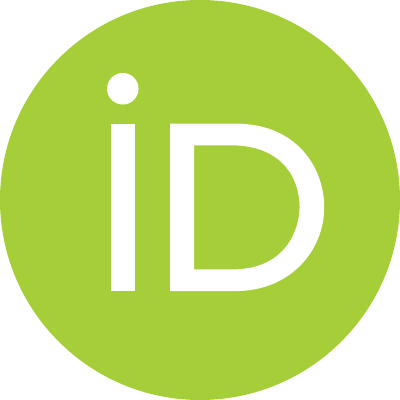}}\\Independent Researcher}

\teaser{
  \includegraphics[width=\columnwidth]{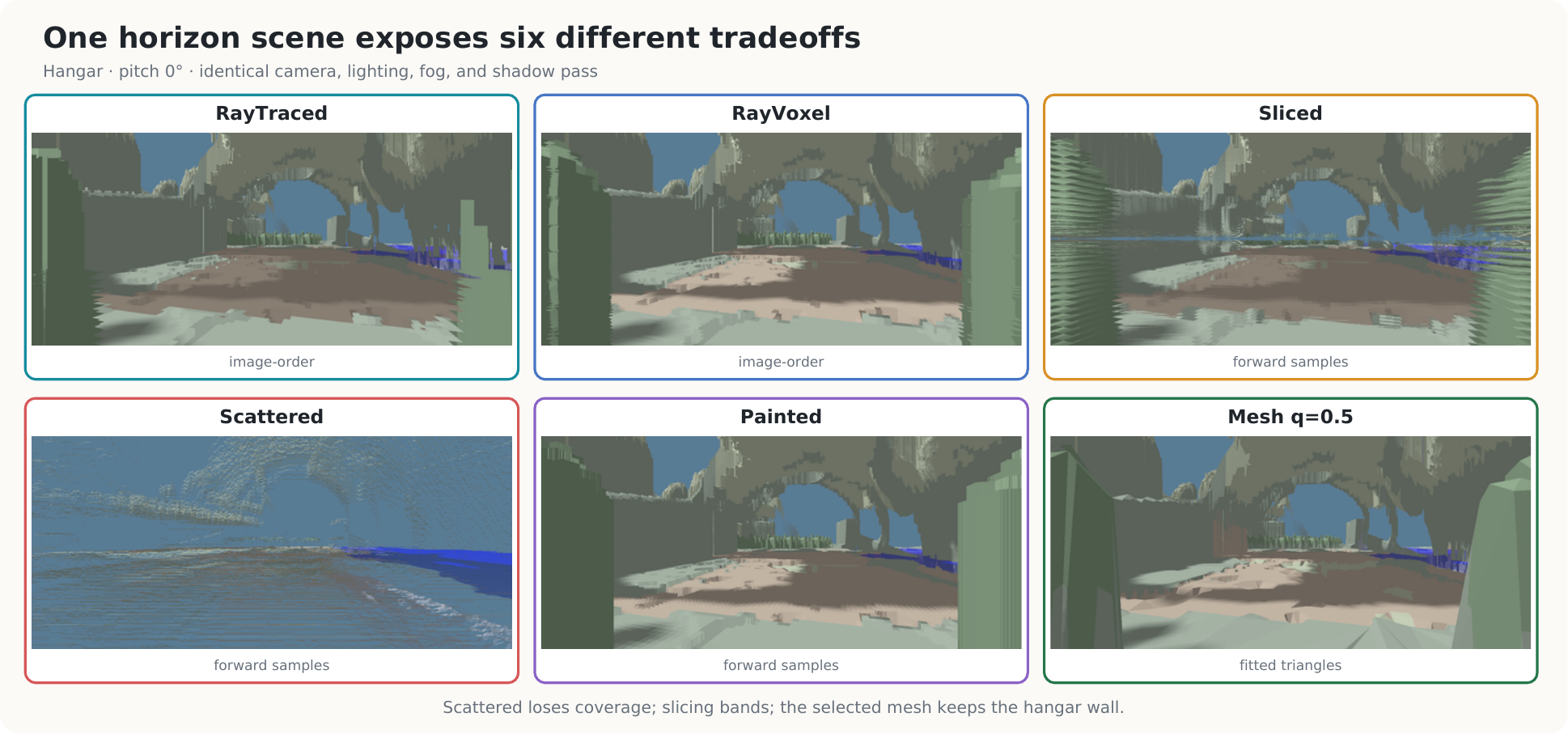}
  \caption{The six selected methods at the hangar horizon scene. Scattered loses coverage; slicing shows grazing bands; the selected mesh keeps the wall.}
  \label{fig:teaser}
}

\maketitle

\begin{abstract}
\small
Terrain level-of-detail is measured almost exclusively on digital elevation models: single-valued, smooth at the sampling scale, sampled from real topography. Game terrain is often none of these. We compare six rendering methods --- height-field ray marching, voxel-accelerated ray marching, sliced proxy geometry, per-sample bar rasterization, compute scattering, and a fitted triangle mesh --- implemented in a single engine over a single data path, on the hand-authored multi-layer terrain of \emph{Vangers} (1998), scored against a CPU ray cast of the same source data. Every method must preserve the two solid intervals available at a ground sample, render at interactive rates, and reflect local terrain destruction without reloading the level. These constraints rule out treating caves as decoration or amortising a static preprocessing step over an immutable map.

From the original game\textquotesingle s top-down camera the six methods look interchangeable. At eye-level horizons they do not: point scattering loses coverage, slicing bands, and an over-simplified mesh can miss a wall. At the selected quality settings a greedy triangulated irregular network (TIN) has the lowest mean frame time on every device we measured, but the fit cost is set by the second layer rather than by floor relief, and making that mesh editable retains 319 MiB of GPU geometry and 535 MiB of CPU triangulation. All six implementations use the same native wgpu / WebGPU API and canonical WGSL. We release the engine, the harness, and a one-command measurement protocol.
\end{abstract}

\section{Introduction}\label{sec:introduction}\label{introduction}

Terrain rendering research usually starts with a digital elevation model: a single-valued surface whose resolution can be traded against screen-space error. Authored game terrain can violate each part of that model. It may be quantised, intentionally discontinuous, and multi-layered, and it is judged from cameras that the original authoring tool never showed. Those differences make familiar reduction ratios and quality settings poor predictors until they are measured on the actual data.

The historical baseline deserves emphasis. Almost three decades ago, \emph{Vangers} rendered this destructible, multi-layer world in software on consumer CPUs while fitting the game and its streamed terrain into a 16 MB-era memory budget \citep{vangerssource}. It succeeded by co-designing the encoding, renderer, and modestly angled oblique top-down views. Its software renderer used a painter\textquotesingle s algorithm, traversing terrain line by line from back to front; that object-order projection later inspired the Scattered method (Section~\ref{sec:scattered}). The distance since then is simultaneously large and small: modern GPUs make arbitrary cameras, common shadowing, and portable programmable shading practical, yet the horizon and cave cases below still punish a representation that is not matched to the data.

\pandocbounded{\includegraphics[width=\columnwidth]{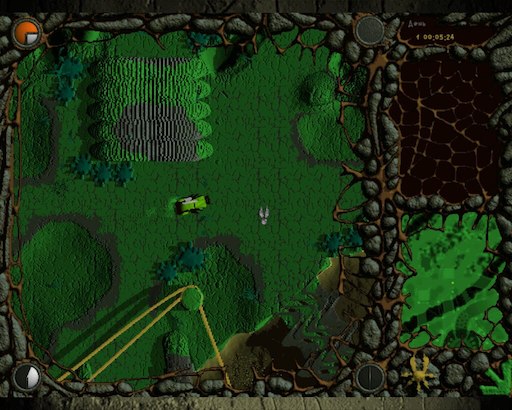}}

This study has three non-negotiable system constraints. First, the renderer must preserve multiple vertical solid intervals, including the underside of an upper slab. Second, it must run inside an interactive frame budget rather than produce an offline conversion. Third, a bounded edit to height and layer metadata must become visible without a level reload or a full rebuild. The last constraint is inherited from the 1998 engine: deformation and terrain destruction are gameplay operations, not authoring-time exceptions; the source release is the primary implementation record \citep{vangerssource}. We report frame latency rather than assigning a hardware-independent meaning to "real-time", and separate steady-state rendering from post-edit maintenance.

\subsection{Related work and scope}\label{sec:related}\label{related-work-and-scope}

The closest published representation is not a conventional height field. \citet{benes2001layered} store a 2D grid of vertical material intervals and edit it with erosion; this is the same broad height-column/voxel compromise as the shipped Vangers data, though with variable geological strata rather than a fixed pair of solids. Their visualization converts the data to height fields or triangles and does not preserve caves interactively. Grounded Heightmap Trees \citep{alonso2008grounded} go further by attaching locally oriented height maps to represent overhangs, tunnels, and large editing events. They are more general than the fixed vertical Vangers encoding but require a changing tree of local parameterizations. QuadStack \citep{graciano2021quadstack} is the closest direct-rendering system: it run-length encodes vertical stacks, compresses coherent neighbours in a quadtree, and ray casts the compressed layers on the GPU. Its published construction and evaluation target compressed volumes and provide no local edit algorithm; modifying a compressed region can require recalculating the representation. Our source is already compact and gameplay edits are required, so we retain direct random access and update only method-specific dirty regions.

Layered Depth Images also store several samples along a ray \citep{shade1998ldi}, but those samples belong to one input camera and are splatted into nearby views. Vangers layers instead live in world-space vertical columns; they survive arbitrary camera motion and are the editable simulation state. The similarity is therefore depth complexity, not representation semantics.

The rendering families themselves are established. Relief mapping uses a linear search followed by refinement through a single height texture \citep{policarpo2005relief}, and GPU landscape ray casting makes the cost output-sensitive \citep{mantler2006landscape}. Maximum mipmaps add conservative hierarchical skipping while remaining cheap enough to update for dynamic, single-valued height fields \citep{tevs2008maximum}. Our first marcher keeps the linear/refinement structure but evaluates two solid intervals; our voxel path generalises the hierarchy to 3D occupancy so it can skip empty caves as well as air. This is deliberately simpler and more updateable than a compressed sparse voxel scene \citep{laine2010svo}.

The three forward methods likewise have clear precedents. Texture-based volume rendering accumulates view-aligned slices \citep{cabral1994volume}, while volume and surface splatting spread each sample over a reconstruction footprint \citep{westover1990footprint,zwicker2001splatting}. Our slicer uses horizontal planes because membership in a vertical interval is then a direct texture test; the resulting grazing-angle bands are the price. Our scatterer writes one nearest-depth pixel per sample rather than a filtered footprint, which explains its holes and temporal speckle. A multi-pixel splat is promising, but it also multiplies contended atomic writes and is left as a measured follow-up rather than folded into this comparison without tuning.

Finally, triangulated irregular network (TIN) fitting and terrain LOD provide the mesh lineage \citep{fowler1979tin,garland1995terrain,duchaineau1997roam,losasso2004clipmaps}. Those systems approximate a single-valued surface. Our mesh shares one topology across three discontinuous altitude fields and locally refits dirty chunks after edits, which is the source of both its unusual fit cost and its update burden.

The implementations in this repository were developed from first principles before this literature audit. That history is not an algorithmic priority claim. The contribution is the controlled comparison under the three constraints above, plus the resulting failure analysis, rather than a claim that ray marching, interval terrain, splatting, slicing, voxels, or greedy TINs are new.

Contributions:

\begin{enumerate}
\def\labelenumi{\arabic{enumi}.}
\tightlist
\item
  A controlled comparison of six terrain rendering methods sharing one engine, one data path, one camera and one fragment-stage shading path, so differences are attributable to the method rather than the surrounding system.
\item
  An evaluation methodology scoring against a CPU ray cast of the source data, decomposed into coverage, geometric and coherence error, with the failure mode of the naive version documented.
\item
  The engine, the harness and the measurement protocol, released under a permissive license (see \emph{Data availability} below).
\item
  A ten-world survey isolating what actually drives fit cost, and the mechanism behind it.
\item
  An edit-path audit showing which methods consume the live interval field directly and which must maintain derived acceleration or mesh data.
\item
  A five-device portability check across Vulkan and Metal using the same validated WGSL shaders and renderer configuration.
\end{enumerate}

\subsection{A decade-long WebGPU testbed}\label{sec:webgpu}\label{a-decade-long-webgpu-testbed}

This comparison is also the record of a long-running implementation, not six algorithms written for one benchmark. vange-rs began in June 2016 and passed its tenth anniversary while this study was being prepared. The basic ray path dates to that first month; the move to the pre-release native wgpu stack and the sliced and scattered paths followed in 2019, Painted in 2020, the complete WGSL migration in 2021, RayVoxel in 2022, and the current fitted mesh in 2026. The techniques therefore accumulated gradually as the engine, API, shader language, and available hardware matured.

The comparison vehicle is the \emph{native} WebGPU API as implemented by wgpu: one validated WGSL source, no backend-specific shaders, no unchecked native commands. Firefox uses wgpu-core to validate WebGPU operations and route them through native graphics APIs, while Naga validates and translates WGSL \citep{malyshau2020webgpu,wgpu,w3cwebgpu,w3cwgsl}. vange-rs moved to wgpu 0.2 in March 2019 and migrated all terrain, object, and debug shaders to WGSL in 2021 \citep{malyshau2021rust}. The same source can be compiled to the web; that is a path-to-the-web proof, not a second evaluation. The numbers in this paper are native Vulkan and Metal runs of that API.

No method bypasses the stack. The comparison therefore asks what quality and performance these unusual terrain techniques achieve \emph{inside} WebGPU\textquotesingle s portable, validated programming model, rather than how far one backend can be special-cased. Agreement of the five result grids across four Vulkan adapters and Apple Metal is the direct portability result; timing remains backend- and device-specific, and Section~\ref{sec:timing} keeps unlike timing mechanisms separate.

A browser smoke test builds the same source for \passthrough{\lstinline!wasm32-unknown-unknown!} and runs the WebGPU-only voxel route in Firefox 152. On a Radeon 890M, Firefox selected its Vulkan WebGPU backend, accepted the canonical WGSL, loaded Fostral, and rendered a 1280$\times$714 canvas. This establishes that a web path exists and looks right; it is not a browser performance result. Commands, versions, and artifact hashes are in \passthrough{\lstinline!paper/browser-smoke.md!}.

\textbf{Data availability.} The engine and evaluation tools are Apache-2.0. The five-device publication batch is pinned at \href{https://github.com/kvark/vange-rs/tree/terrain-paper}{\passthrough{\lstinline!terrain-paper!}}. Fostral world data is published by Association K-D Lab under \href{https://creativecommons.org/licenses/by-sa/4.0/}{CC BY-SA 4.0}; the canonical tree is \href{https://github.com/KranX/Vangers/tree/master/data/thechain/fostral}{\passthrough{\lstinline!KranX/Vangers!} \passthrough{\lstinline!data/thechain/fostral!}} at commit \passthrough{\lstinline!f1ad7d7!}. The harness fetches that commit; we do not redistribute a second archive. The ten-world fit survey also uses nine other shipped levels that are not in that grant --- those rows require a lawfully obtained game copy. Derived figures and the supplemental video use Fostral and carry the same attribution.

\section{The Data}\label{sec:data}\label{the-data}

Vangers terrain is a height map with a per-texel-\emph{pair} dual encoding. A texel is either single-valued, or carries three altitudes: a floor (\passthrough{\lstinline!low!}), a cave ceiling (\passthrough{\lstinline!mid!}) and a slab top (\passthrough{\lstinline!high!}), with the even texel of a pair storing \passthrough{\lstinline!low!}, the odd one \passthrough{\lstinline!high!}, and \passthrough{\lstinline!mid!} derived from delta bits in both. Fostral, the level used throughout, is 2048$\times$16384 and 10.9\% double-level.

\pandocbounded{\includegraphics[width=\columnwidth]{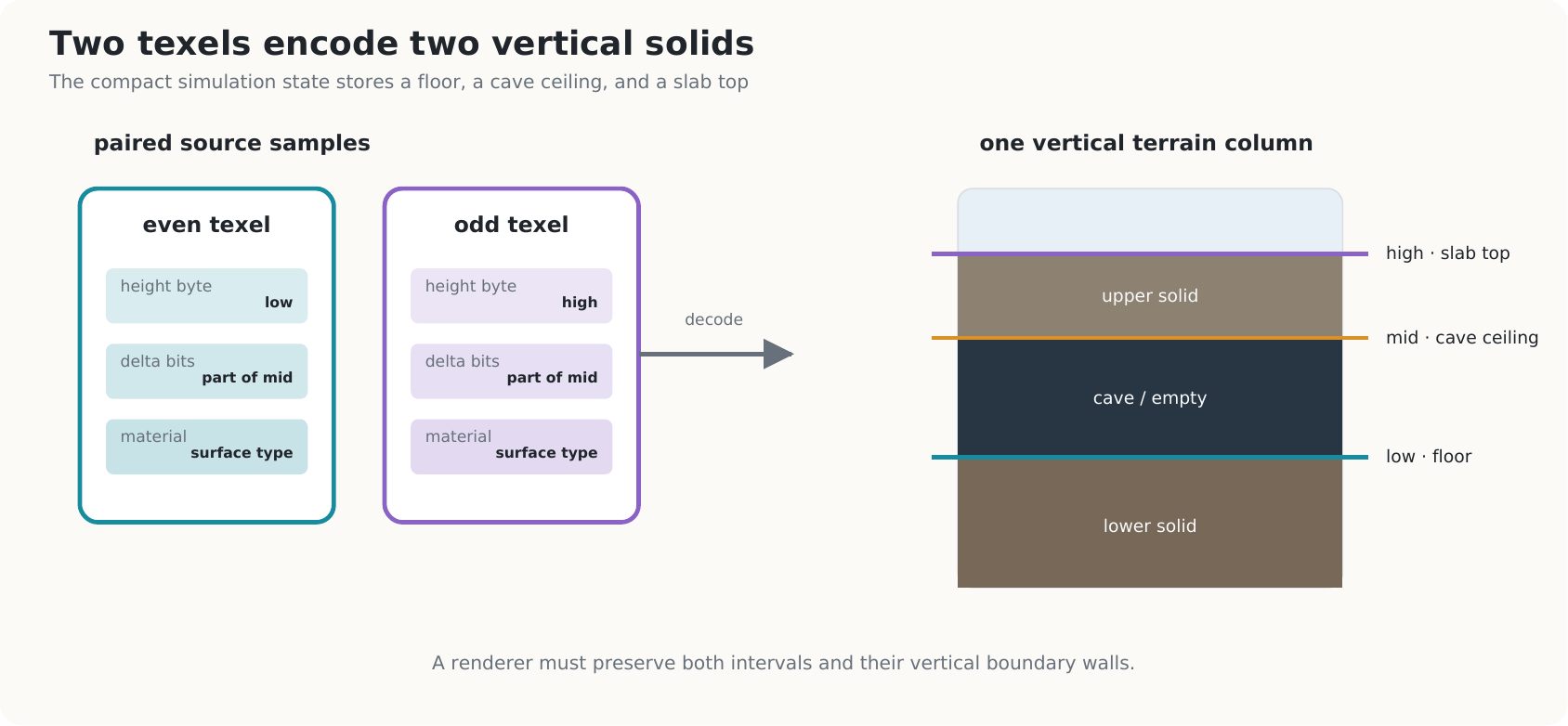}}

Two properties matter for everything that follows. The surface is \textbf{authored, not sampled}: cliffs are vertical by construction rather than by steep gradient, and adjacent texels routinely differ by tens of units. And \passthrough{\lstinline!mid!}/\passthrough{\lstinline!high!} are \textbf{structural, not fields}: they describe where a slab is, and interpolating them across the region boundary produces geometry that exists nowhere. Section 6.3 is about what that costs.

The height and metadata arrays are also mutable simulation state. A local gameplay event may change both altitude and whether the upper interval exists. The renderer receives a dirty rectangle, uploads the modified rows, and must make that edit visible. Ray, Sliced, Painted, and Scattered read the updated texture directly. RayVoxel incrementally rebuilds affected occupancy cells and their ancestors; Mesh locally refines affected chunks and replaces their GPU buffers. Thus all six support edits, but their time-to-consistency and update cost are different quantities that steady-state frame time does not capture. Section~\ref{sec:edit} shows the same crater before and after on a direct reader, an incremental occupancy rebuild, and a local mesh refit, and reports both CPU submit-and-wait and GPU draw cost for the first updated frame.

\section{Methods Compared}\label{sec:methods}\label{methods-compared}

All methods use the same decoded height and material data, camera, palette, fog, diffuse-lighting function, and shadow map. The method under test is the way visible surface samples reach the framebuffer. The mesh alone supplies a geometric normal; the other five estimate a height-field gradient in the shared shading function.

Scattered stores 24 bits of depth and 8 bits of terrain type in its intermediate buffer, then reconstructs world position and applies the same diffuse and shadow evaluation in its resolve pass as the other five methods.

The methods fall into three groups. The two image-order methods cast one ray per pixel. Sliced, Painted, and Scattered enumerate samples of the encoded volume and project them forward. Mesh performs an offline fit and rasterizes the resulting explicit surface.

{\small
\begin{tabularx}{\columnwidth}{@{}l l Y Y@{}}
\toprule
method & order & primitive & edit response \\
\midrule
Height-field ray march & image & per-pixel ray & direct texture read \\
Voxel-accelerated ray march & image & per-pixel ray & incremental occupancy rebuild \\
Sliced & object & horizontal proxy quads & direct texture read \\
Painted & object & bars per ground sample & direct texture read \\
Scattered & object & compute-scattered points & direct texture read \\
\textbf{Mesh (TIN)} & object & fitted triangles & local chunk refit \\
\bottomrule
\end{tabularx}
}

Every row evaluates both encoded solid intervals. "Direct" means the next draw after the dirty texture upload sees the edit; it does not mean the CPU upload itself is free. Derived rows can remain interactive only if the dirty work is bounded or spread across frames.

\subsection{Height-field ray march}\label{sec:ray}\label{height-field-ray-march}

\pandocbounded{\includegraphics[width=\columnwidth]{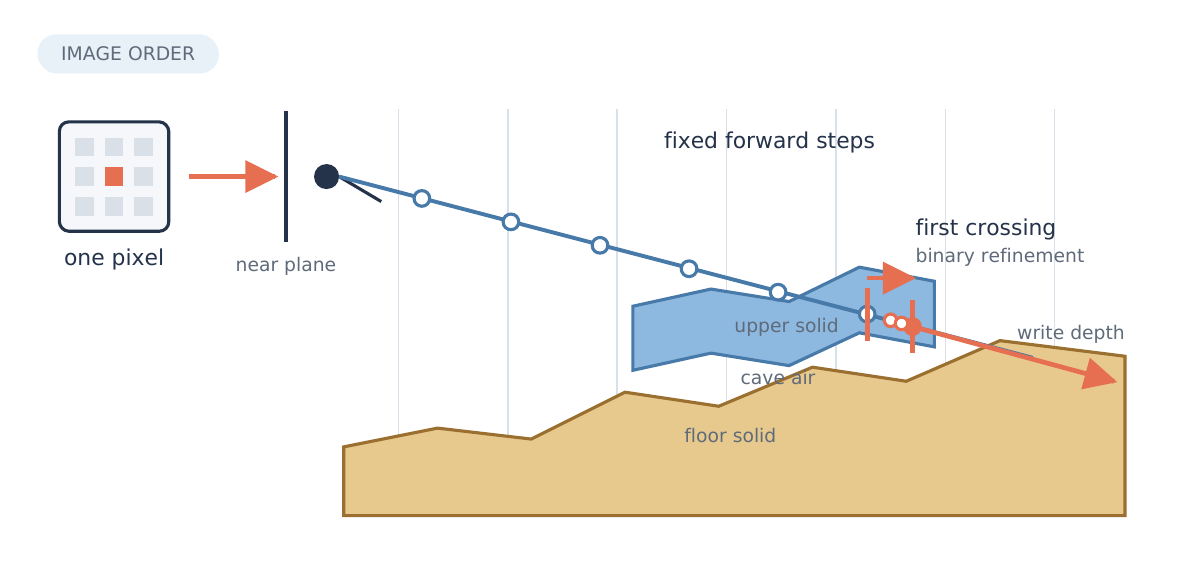}}

A full-screen pass reconstructs the near and far world-space point for each pixel. The segment is sampled uniformly until it enters the encoded solid, then four binary-search iterations refine the first crossing. For a single-level texel, points at or below \passthrough{\lstinline!high!} are solid. For a dual-level texel, \passthrough{\lstinline!z <= low!} and \passthrough{\lstinline!mid <= z <= high!} are solid. Testing this predicate directly is important: it detects floors, slab tops, and cave ceilings for rays travelling in either vertical direction.

\begin{lstlisting}[float=false,language=]
function RAYMARCH(o, d, steps):
    a, b $\leftarrow$ o, clip(o, d)                 // far plane or z = 0
    for i $\leftarrow$ 1 to steps:
        c $\leftarrow$ a + (b -- a) / (steps + 1)
        if SOLID(c): b $\leftarrow$ c; break
        else a $\leftarrow$ c
    for i $\leftarrow$ 1 to 4:                      // bisection
        c $\leftarrow$ (a + b) / 2
        if SOLID(c): b $\leftarrow$ c else a $\leftarrow$ c
    return b if hit else MISS

SOLID(p) := p.z $\leq$ low(p.xy)  $\lor$  mid(p.xy) $\leq$ p.z $\leq$ high(p.xy)
\end{lstlisting}

The publication setting uses 128 forward samples over the clipped ray segment. The budget is exposed as \passthrough{\lstinline!--ray-steps!} and is included in the uniform tuning sweep. A sample interval can still skip a thinner feature; that is the method\textquotesingle s characteristic quality/performance tradeoff. Misses return the cleared far depth instead of manufacturing a hit at the ground plane. Hits write the reconstructed depth and therefore compose normally with rasterized objects. The method needs no preprocessing or auxiliary storage and remains the WebGL2 fallback.

\subsection{Voxel-accelerated ray march}\label{sec:voxel}\label{voxel-accelerated-ray-march}

\pandocbounded{\includegraphics[width=\columnwidth]{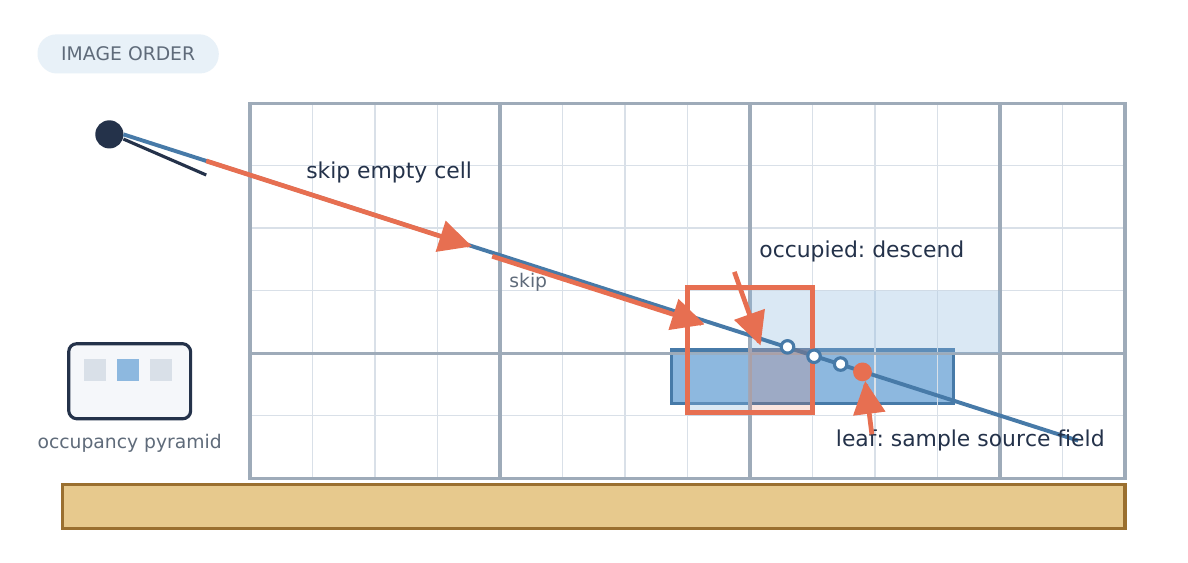}}

This path accelerates the same per-pixel query with a conservative occupancy pyramid. The finest level stores one occupancy bit per voxel in Morton-coded 8$\times$8$\times$8 tiles; each higher level is the union of eight children. Traversal uses a hierarchical DDA: an empty cell advances directly to its exit plane, while an occupied cell descends until the leaf level, where fixed sampling tests the original height data. The structure skips only known-empty space; it does not replace the source surface with cubes.

\begin{lstlisting}[float=false,language=]
function RAYVOXEL(o, d):
    lod $\leftarrow$ coarsest
    while outer steps remain:
        if occupied(cell, lod) and lod > 0:
            lod $\leftarrow$ lod -- 1; continue      // descend into a child
        advance o to the cell exit
        if occupied and lod = 0:
            if a linear sample in the cell is SOLID: return hit
        else if the exit allows it: lod $\leftarrow$ lod + 1
    return MISS
\end{lstlisting}

The publication comparison uses the coarse \passthrough{\lstinline!(4,8,2)!} occupancy grid (18.29 MiB). The renderer\textquotesingle s shipping \passthrough{\lstinline!(2,4,1)!} grid needs 153 MiB and did not fit the software rasterizer used for tuning; Section~\ref{sec:tuning} records that choice. Every RayVoxel number in this paper is the coarse configuration.

The GPU builds the hierarchy incrementally because a whole-level dispatch can exceed driver watchdog limits. Preparation therefore appears separately in Section~\ref{sec:prep}. Runtime traversal has outer- and inner-step budgets. Dense vertical variation can exhaust them, and the storage-buffer requirement excludes the WebGL2 path.

\subsection{Sliced}\label{sec:sliced}\label{sliced}

\pandocbounded{\includegraphics[width=\columnwidth]{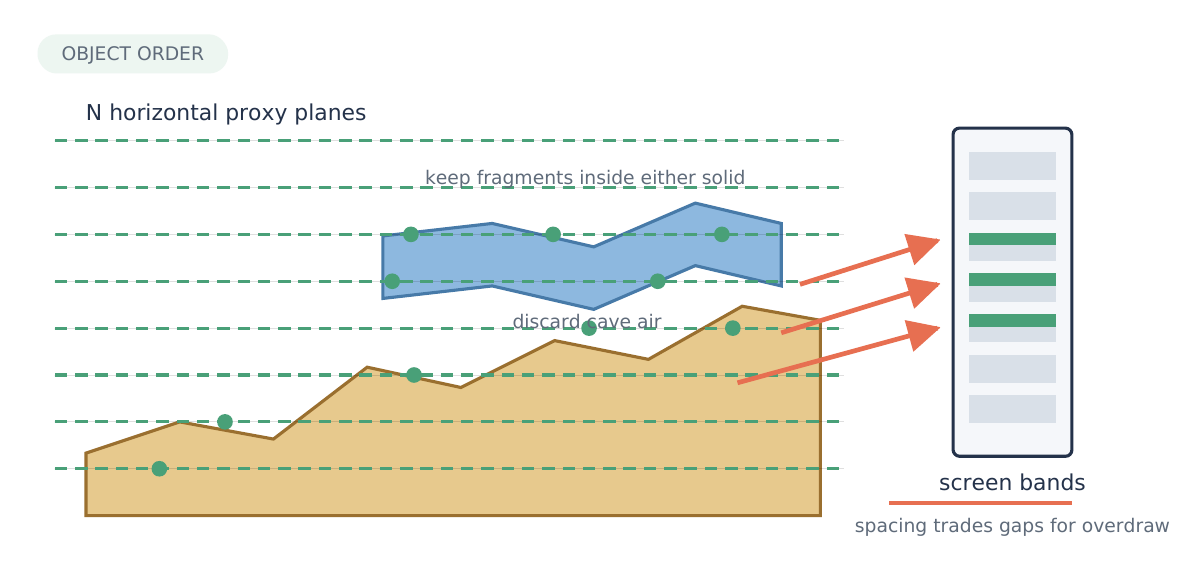}}

The renderer draws \passthrough{\lstinline!N!} horizontal quads across the visible terrain bounds. For each fragment, the surface decoder keeps the sample when its altitude is inside the floor column or upper slab and discards it in empty space. The slices are spread over the full altitude range; reducing \passthrough{\lstinline!N!} therefore coarsens the representation rather than deleting the lowest part of it.

\begin{lstlisting}[float=false,language=]
for k $\leftarrow$ 1 to N:
    z $\leftarrow$ z_max -- k $\cdot$ (z_max / N)
    draw a horizontal quad at z over the camera footprint
    for each fragment p:
        if p.z $\leq$ low(p.xy) or mid $\leq$ p.z $\leq$ high: shade(p)
        else discard
\end{lstlisting}

At the 256 quantized altitude levels, one slice per unit samples every representable height. The tuned setting uses 512 slices. At grazing angles, however, discrete planes remain visible as coherent horizontal bands and can repeatedly cover a silhouette. This pattern can resemble a cast shadow in a comparison image, but it is slice quantization; the shadow lookup is the same one used by every other method.

\subsection{Painted}\label{sec:painted}\label{painted}

\pandocbounded{\includegraphics[width=\columnwidth]{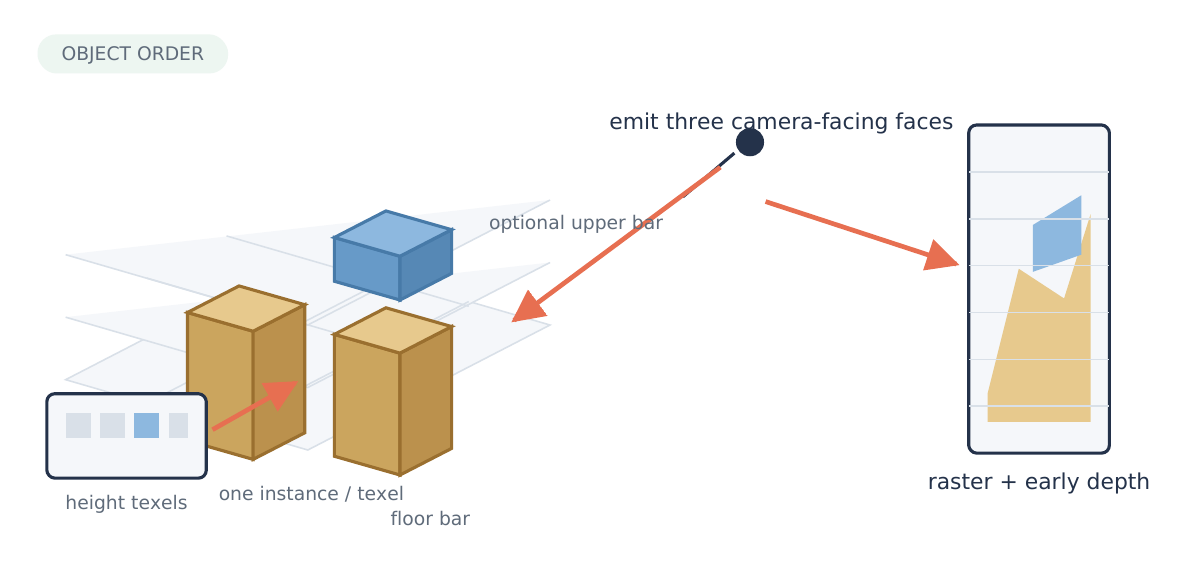}}

Painted turns each ground sample into explicit column geometry. One bar extends from zero to the floor; a dual-level sample adds a bar from cave ceiling to slab top. The vertex shader emits only the three faces oriented toward the camera. It derives every position from the instance index, so no per-column vertex stream is uploaded.

\begin{lstlisting}[float=false,language=]
for each ground sample (x, y) in the camera footprint, front to back:
    emit a bar [0, low] with the three camera-facing faces
    if dual: emit a bar [mid, high]
rasterize with the ordinary depth test
\end{lstlisting}

The instance range is an axis-aligned bound of the camera footprint on the terrain. Ordering the generated samples front-to-back lets early depth tests remove most covered fragments; the original implementation measured 96\% early-Z rejection for its test view. Cost still scales with ground samples inside the footprint rather than with visible pixels, which explains its poor downward-looking timings. Although it also processes terrain in object order, its explicit bar geometry is not a reconstruction of the original software renderer.

\subsection{Scattered}\label{sec:scattered}\label{scattered}

\pandocbounded{\includegraphics[width=\columnwidth]{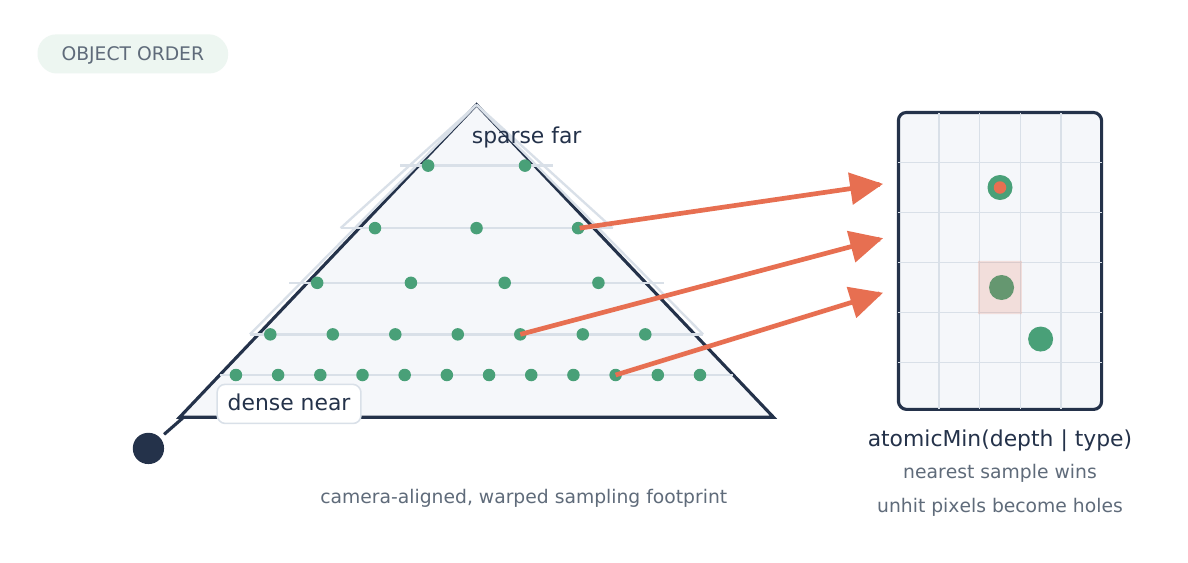}}

Scattered is the method that inherits the original engine\textquotesingle s object-order, line-by-line painter (Section~\ref{sec:introduction}) and recasts it as a parallel atomic scatter. A compute grid samples a camera-aligned ground footprint whose longitudinal coordinate is warped to spend more samples nearby. Each invocation walks both encoded vertical intervals and projects point samples into a screen buffer. Parallel GPU work cannot rely on painter ordering, so a 32-bit \passthrough{\lstinline!atomicMin!} selects the nearest result while retaining 24 bits of depth and 8 bits of terrain type. A full-screen resolve reconstructs world position and performs material, diffuse, fog, cave-ambient, and shadow evaluation.

\begin{lstlisting}[float=false,language=]
clear the screen buffer to $\infty$
for each warped footprint sample (x, y), in parallel:
    scatter (x, y, mix(low, 0, t)) for t $\in$ [0, 1]
    if dual: scatter (x, y, mix(high, mid, t))
scatter(p): atomicMin(buf[project(p)], pack(depth(p), type))
resolve: reconstruct world position from the packed depth and shade
\end{lstlisting}

The density vector controls the compute grid. Unlike rasterization, point projection does not guarantee pixel coverage; undersampling appears as isolated gaps and unstable pixels, especially near the horizon. The coherence metric in Section~\ref{sec:metrics} is intended to expose exactly this failure.

\subsection{Mesh}\label{sec:mesh}\label{mesh}

\pandocbounded{\includegraphics[width=\columnwidth]{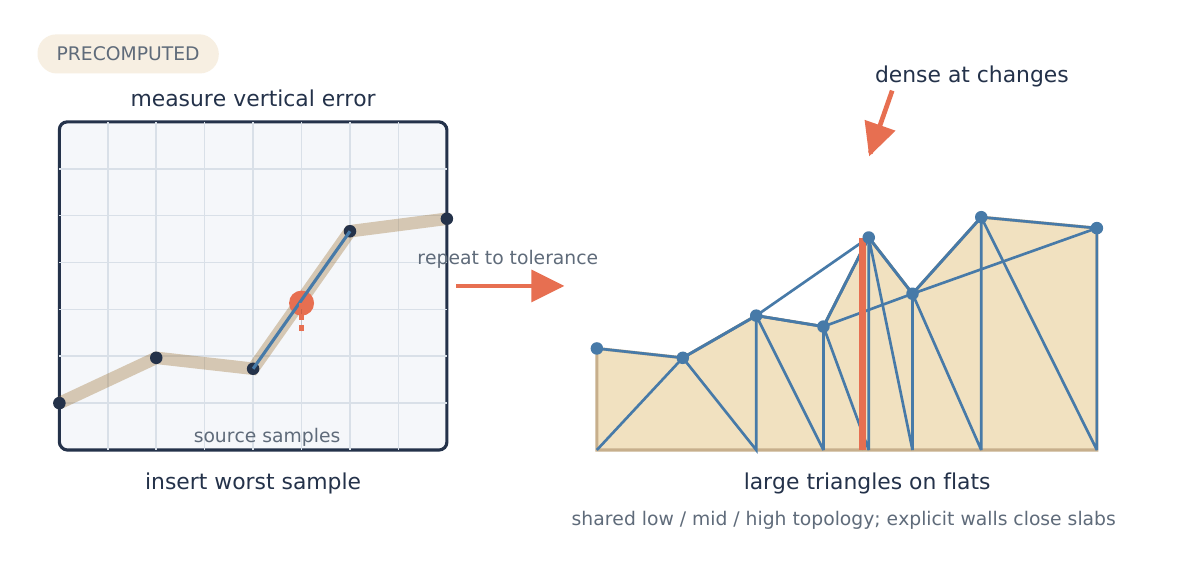}}

The mesh path fits an explicit surface with greedy point insertion following \citet{garland1995terrain}. Each triangle tracks the source sample with the largest vertical error; the globally worst sample is inserted until the tolerance is met. Integer grid coordinates allow orientation and incircle predicates to remain exact on the many cocircular cells.

One planar triangulation is shared by \passthrough{\lstinline!low!}, \passthrough{\lstinline!mid!}, and \passthrough{\lstinline!high!}. A sample is inserted when any of the three fields exceeds tolerance, and boundary walls close the upper slab where dual-level encoding begins or ends. This coupling is the source of the fit-cost result in Section~\ref{sec:findings}.

\begin{lstlisting}[float=false,language=]
for each 128$\times$128 chunk:
    T $\leftarrow$ Delaunay of the four corners
    while max vertical error of T over {low, mid, high} > $\tau$:
        insert the worst sample and restore Delaunay
    emit floor triangles; emit slab + walls on dual-region edges
draw the selected LOD with ordinary rasterization
\end{lstlisting}

The level is partitioned into 128$\times$128 chunks with three fitted detail levels. Chunks are frustum culled and selected by camera distance. Boundaries use a stable finest-level vertex sequence so neighbouring LODs do not crack. The tradeoff moves out of the frame loop: fitting is a blocking CPU cost, the full-level renderer retains all three GPU LODs plus the editable CPU triangulation, and edits require local refinement. At the selected q=0.5 those explicit method-specific allocations are reported in Section~\ref{sec:edit}; they are substantially more than counting only the finest LOD. Once present, the geometry uses the conventional raster pipeline and supplies the triangle\textquotesingle s own geometric normal for lighting, which is defined on vertical walls and cave ceilings where a height-field gradient is not.

The explicit mesh can be reused as a conventional triangle collider, which mainstream rigid-body libraries accept and the original two-interval height map is not. That reuse is a system-level consequence, not a measurement here. It also widens the edit obligation: a refitted visual chunk would need its collider replaced on the same consistency boundary.

\section{Evaluation}\label{sec:eval}\label{evaluation}

\subsection{Reference}\label{sec:reference}\label{reference}

A CPU ray cast of the same height data with the same camera, marching coarsely and bisecting within the bracketing interval. It yields a sky mask and a distance field.

Two mistakes in building this are worth recording because both produced plausible, wrong numbers for some time.

The reference camera\textquotesingle s focal length was scaled by render height while the renderer\textquotesingle s is fixed in pixels. The two agree only at one resolution; at 1080p the reference would have had a 3.6$\times$ wider field of view than the image it was scoring.

The solidity test guarded the floor comparison with \passthrough{\lstinline!z >= 0!}, which makes a texel of height zero unhittable --- no \passthrough{\lstinline!z!} satisfies both \passthrough{\lstinline!z >= 0!} and \passthrough{\lstinline!z < 0!}. Sea level is a real height on these maps, so every downward ray over water was classified as sky. It cost 23\% of a straight-down frame.

\subsection{Metrics}\label{sec:metrics}\label{metrics}

\textbf{Coverage}, decomposed. \passthrough{\lstinline!see-through!} is solid terrain the renderer left as background; \passthrough{\lstinline!covers-sky!} is background it filled in. The decomposition is what makes them diagnostic: only the first moves when a renderer is genuinely missing geometry, and both move together when the reference is the one disagreeing about a silhouette.

\textbf{Geometry.} Median and 95th-percentile distance error where renderer and reference agree something is present. This must be read comparatively: its floor is set by grazing incidence and by the scene\textquotesingle s sampling discontinuities (Section~\ref{sec:offset}).

\textbf{Coherence.} The fraction of pixels whose distance disagrees with their own 3$\times$3 neighbourhood, in excess of the reference doing the same.

Coverage alone cannot detect over-drawing: a renderer that interpolates a slab across region boundaries covers every pixel that should be covered. Comparing depth found that class of defect; coherence finds striping, speckle, and cracks that correct-on-average depth still misses.

\subsection{Protocol}\label{sec:protocol}\label{protocol}

Every number in this paper is produced by one command with no arguments (\passthrough{\lstinline!tools/compare-terrain.py!}), whose defaults \emph{are} the publication configuration: every method, twelve viewpoints (three per pitch, at 0°, --30°, --60°, --90°), 1280$\times$800, 40 timed frames after per-method warmup. The harness builds the binaries, fetches and converts the level on first run, and names its output after the adapter wgpu selected, so runs collected from different machines merge without hand-labelling (\passthrough{\lstinline!tools/merge-bench.py!}). The design goal is that a measurement session on a new device costs its owner one invocation and about an hour --- the protocol is only as reproducible as it is cheap to follow.

On Vulkan, frame times come from GPU timestamp queries around the frame\textquotesingle s command encoder. Encoder-level timestamps did not reliably bracket this multipass workload on Metal: the returned intervals were quantised, nearly method-independent, and much shorter than submit-to-completion. This is a measurement failure, not evidence that Metal rendered incorrectly. wgpu explicitly does not guarantee strict ordering for arbitrary command-encoder timestamps, and its Metal backend may defer such a write to the next native pass. We treated the pair as a guaranteed enclosing interval, which the API does not promise. A future GPU-only measurement must timestamp each render and compute pass and sum those intervals; the present Apple run conservatively uses the retained CPU submit-and-wait average instead. Each row records its timing source because the two mechanisms are not directly comparable (Section~\ref{sec:timing}). The publication configuration renders the same 1024² height-field shadow map for every method. This deliberately measures a complete, visually comparable terrain frame rather than an isolated discovery pass. The JSON records the shadow mode, and \passthrough{\lstinline!--no-shadows!} is available for a separate method-only diagnostic run. The final hardware batch uses this protocol; visual review of all five grids confirms that the common pass reaches every column. One-time costs are recorded separately as setup / first frame / warmup (Section~\ref{sec:prep}), since per-frame figures structurally exclude them. Accuracy is expected to be device-independent; the merge tool reports a baseline once and cross-checks every field from the other devices, because an adapter that disagrees about geometry is a finding, not noise.

\subsection{Dynamic-edit protocol}\label{sec:edit-protocol}\label{dynamic-edit-protocol}

The headless runner\textquotesingle s \passthrough{\lstinline!--dig!} path removes altitude and upper-layer metadata inside a radius-48 crater centred at (1024, 8192) and submits the same dirty rectangle used by the interactive game. The timing camera is at (1024, 8350), 40 units above the original surface, yaw 180°, pitch --30°. For each method the protocol records (1) the median of five independently constructed first post-edit frames, both as CPU submit-and-wait and as GPU timestamps when the adapter provides them, (2) consistency after 1, 2, 4, 8, or 16 frames, and (3) colour and depth agreement with a fresh renderer constructed from the already-edited level. Consistency means at most 0.01\% hit/miss disagreement and p95 depth difference at most 0.1 world unit. The Section~\ref{sec:edit} figure uses a closer, higher view of the same crater so the bowl is readable; the numbers always come from the timing camera.

The same run reports explicit persistent method data: renderer-owned GPU buffers and retained CPU acceleration or fitting structures. Shared terrain, palette, color/depth and shadow textures, pipelines, staging allocations, and opaque driver memory are excluded. This narrower measure is portable and auditable through wgpu; whole-process RSS or backend heap telemetry is not. The edit test remains separate from the steady-state protocol because averaging an update into 40 ordinary frames would hide the cost that the constraint exists to expose.

\section{Results}\label{sec:results}\label{results}

The publication comparison uses one six-method set throughout: RayTraced 128, RayVoxel 100, Sliced 512, Scattered 4,4,4, Painted, and Mesh q=0.5. Every run is 1280$\times$800 with far distance 600, 40 timed frames, a common 1024² ray-traced shadow pass, and the same twelve camera records. The five-device batch is AMD Radeon 780M, AMD Radeon RX 7900 XT, Intel RPL-U, NVIDIA GeForce RTX 5070, and Apple M3. Vulkan rows use GPU timestamps; the M3 uses CPU submit-and-wait. Complete per-view tables and driver versions are in \passthrough{\lstinline!paper/results.md!}.

\subsection{Pitch is the axis that separates them}\label{sec:pitch}\label{pitch-is-the-axis-that-separates-them}

Values below are arithmetic means over the three views at each pitch, reported as see-through / coherence error (\%), from the Radeon 780M run and cross-checked on the other four adapters. \passthrough{\lstinline!covers-sky!} is omitted because its pitch mean is at most 0.2\% for every selected configuration (Section~\ref{sec:offset}).

\pandocbounded{\includegraphics[width=\columnwidth]{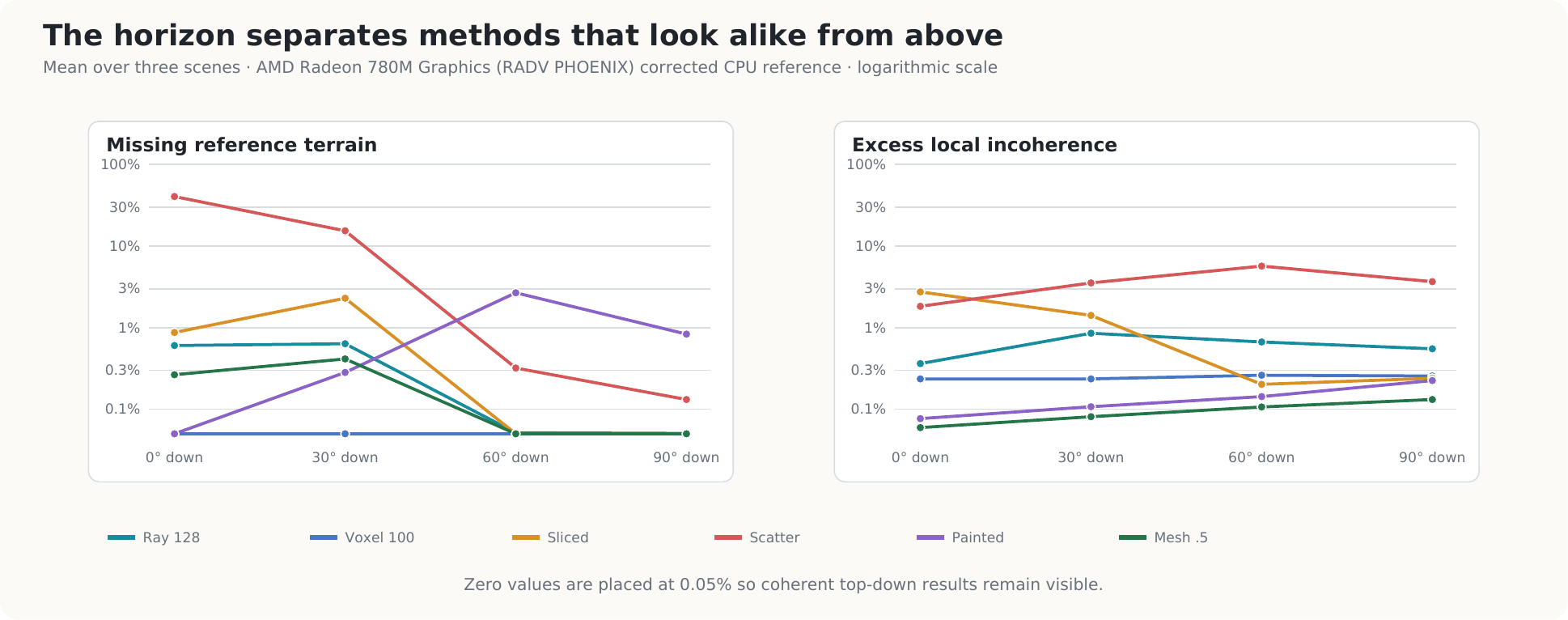}}

{\footnotesize\setlength{\tabcolsep}{3.5pt}
\begin{tabular}{@{}lrrrrrr@{}}
\toprule
pitch & Ray 128 & Voxel 100 & Sliced & Scattered & Painted & Mesh .5 \\
\midrule
0° & 0.6 / 0.4 & 0.0 / 0.2 & 0.9 / 2.7 & \textbf{40.4 / 1.8} & 0.0 / 0.1 & 0.3 / 0.1 \\
--30° & 0.6 / 0.9 & 0.0 / 0.2 & 2.3 / 1.4 & \textbf{15.4 / 3.5} & 0.3 / 0.1 & 0.4 / 0.1 \\
--60° & 0.1 / 0.7 & 0.0 / 0.3 & 0.0 / 0.2 & 0.3 / \textbf{5.7} & 2.7 / 0.1 & 0.0 / 0.1 \\
--90° & 0.1 / 0.5 & 0.0 / 0.3 & 0.0 / 0.2 & 0.1 / \textbf{3.7} & 0.8 / 0.2 & 0.0 / 0.1 \\
\bottomrule
\end{tabular}
}

The image-order methods join the coherent group. Scattering is the horizon outlier: its three scenes leave 19.1--73.2\% of reference terrain uncovered, while the other methods\textquotesingle{} pitch means lie between 0.0\% and 0.9\%. Looking down removes its coverage deficit but not its point-scale incoherence. Slicing shows the complementary signature: good coverage but visible horizontal bands, measured as 2.7\% coherence error at 0°.

The selected mesh sits with the coherent group. At the hangar view q=0.0 left 11.0\% uncovered; q=0.5 leaves 0.5\% and recovers the wall. The five image grids agree on that geometry: hangar Mesh see-through is 0.519\% on every adapter. The only material cross-device spread is inside the already failing Scattered hangar row (73.2\% on the 780M versus 74.0\% on the M3).

\textbf{Methods developed from top-down screenshots can conceal failure modes that become dominant at eye level.} Point scattering loses coverage, slicing bands, and an aggressively simplified mesh can miss a scene-specific wall; the image-order methods, Painted, and the selected mesh remain mutually coherent.

\subsection{Frame time}\label{sec:timing}\label{frame-time}

The timing table uses the same six configurations as Section~\ref{sec:pitch}. Vulkan rows use GPU timestamps; Apple M3 rows marked \passthrough{\lstinline!*!} use CPU submit-and-wait because encoder-level timestamps do not enclose the Metal workload (Section~\ref{sec:protocol}).

\pandocbounded{\includegraphics[width=\columnwidth]{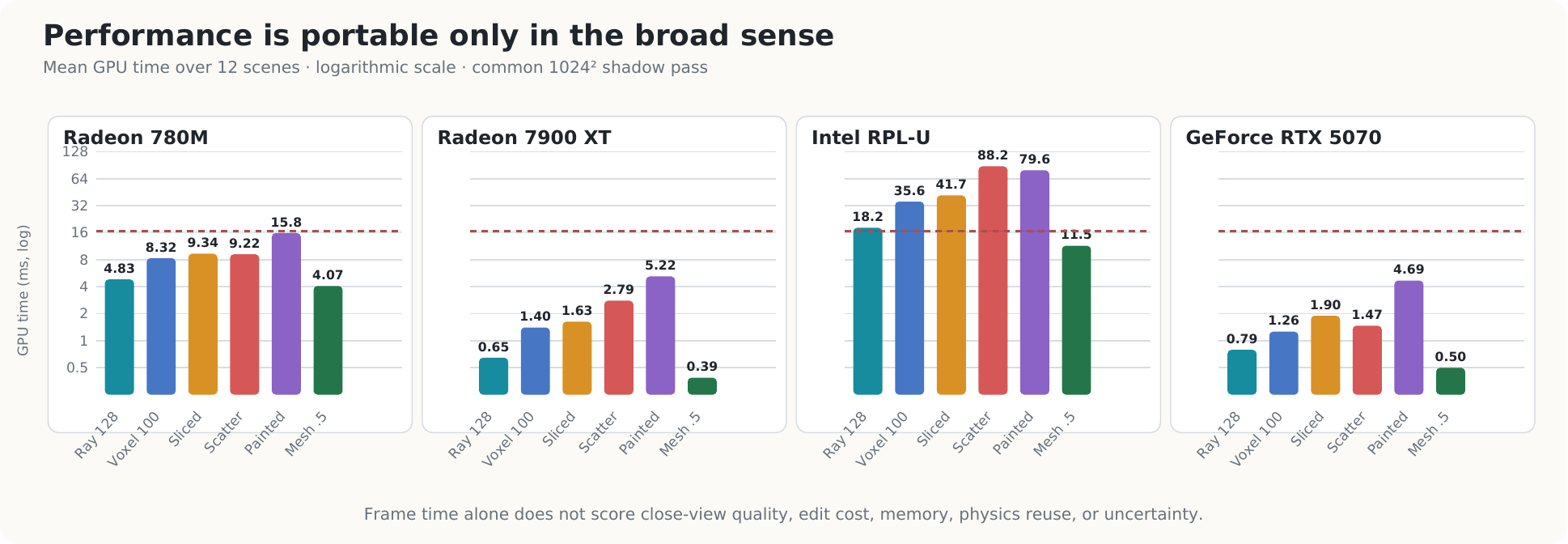}}

{\footnotesize\setlength{\tabcolsep}{3.2pt}
\begin{tabular}{@{}lrrrrrr@{}}
\toprule
device / pitch & Ray 128 & Voxel & Sliced & Scat.$^\dagger$ & Painted & Mesh .5 \\
\midrule
780M / 0° & 4.562 & 8.143 & 10.437 & 14.179 & 8.441 & \textbf{4.072} \\
780M / --30° & 4.871 & 9.123 & 7.809 & 8.658 & 12.853 & \textbf{4.152} \\
780M / --60° & 4.838 & 8.254 & 8.974 & 6.970 & 20.058 & \textbf{3.974} \\
780M / --90° & 5.044 & 7.762 & 10.142 & 7.072 & 22.005 & \textbf{4.081} \\
7900 XT / 0° & 0.574 & 1.365 & 1.781 & 7.897 & 3.035 & \textbf{0.384} \\
7900 XT / --30° & 0.660 & 1.522 & 1.343 & 1.315 & 4.658 & \textbf{0.395} \\
7900 XT / --60° & 0.658 & 1.415 & 1.604 & 0.991 & 6.837 & \textbf{0.385} \\
7900 XT / --90° & 0.693 & 1.308 & 1.790 & 0.974 & 6.332 & \textbf{0.383} \\
Intel / 0° & 15.751 & 34.681 & 54.498 & 104.370 & 45.628 & \textbf{11.745} \\
Intel / --30° & 18.106 & 39.596 & 34.730 & 89.523 & 62.300 & \textbf{11.426} \\
Intel / --60° & 17.591 & 34.433 & 34.862 & 87.927 & 86.566 & \textbf{11.256} \\
Intel / --90° & 21.227 & 33.561 & 42.657 & 70.993 & 124.020 & \textbf{11.388} \\
RTX 5070 / 0° & 0.722 & 1.251 & 2.259 & 1.721 & 3.258 & \textbf{0.503} \\
RTX 5070 / --30° & 0.809 & 1.419 & 1.544 & 1.532 & 4.155 & \textbf{0.501} \\
RTX 5070 / --60° & 0.782 & 1.226 & 1.774 & 1.327 & 5.676 & \textbf{0.493} \\
RTX 5070 / --90° & 0.845 & 1.146 & 2.018 & 1.304 & 5.679 & \textbf{0.496} \\
M3* / 0° & 8.219 & 12.378 & 12.284 & 40.524 & 16.944 & \textbf{5.307} \\
M3* / --30° & 8.057 & 13.551 & 11.017 & 11.443 & 32.664 & \textbf{5.936} \\
M3* / --60° & 8.172 & 12.173 & 12.110 & 10.681 & 49.750 & \textbf{6.302} \\
M3* / --90° & 8.915 & 11.862 & 13.124 & 11.161 & 31.530 & \textbf{6.135} \\
\bottomrule
\end{tabular}
}

\textbf{* Apple M3 values are CPU submit-and-wait means.} They include command submission and the completion round trip, so they support comparisons among methods on that device but not absolute comparisons with the Vulkan GPU rows.

At the selected quality point Mesh q=0.5 has the lowest twelve-scene mean on every adapter: 4.070 ms on the 780M, 0.387 ms on the 7900 XT, 11.454 ms on Intel, 0.498 ms on the RTX 5070, and 5.920 ms CPU on the M3. RayTraced 128 is second on four of the five; the 7900 XT and 5070 put the two within 0.3 ms, while Intel and the M3 open a larger gap. RayVoxel is consistently about 1.6--2$\times$ RayTraced. Portability of an API and shader does not imply portability of a performance ranking among the slower methods, but it does not disturb this mesh-versus-ray order.

Painted has the clearest orientation dependence on every adapter, becoming progressively slower as more ground samples enter its emitted footprint. Scattered is less orderly: the hangar takes 29.3 ms on the 780M, 21.8 ms on the 7900 XT, 148.9 ms on Intel, and 99.8 ms CPU time on Apple, but only 2.5 ms on NVIDIA. $^\dagger$The pitch-0 arithmetic mean is therefore dominated by a scene-and-adapter interaction and is not a general horizon cost. On Intel the derived-volume and forward methods cost 34--124 ms while RayTraced and Mesh stay near 11--21 ms.

Within-session 95\% intervals from the 40 frame samples, over the twelve-scene mean:

{\footnotesize\setlength{\tabcolsep}{3pt}
\begin{tabular}{@{}lrrrrrr@{}}
\toprule
device & Ray 128 & Voxel & Sliced & Scat. & Paint & Mesh \\
\midrule
780M & 4.83$\pm$0.12 & 8.32$\pm$0.01 & 9.34$\pm$0.12 & 9.22$\pm$0.04 & 15.84$\pm$0.09 & 4.07$\pm$0.05 \\
7900 XT & 0.65$\pm$0.01 & 1.40$\pm$0.00 & 1.63$\pm$0.01 & 2.79$\pm$0.00 & 5.22$\pm$0.00 & 0.39$\pm$0.00 \\
Intel & 18.17$\pm$0.10 & 35.57$\pm$0.06 & 41.69$\pm$0.42 & 88.20$\pm$0.16 & 79.63$\pm$0.44 & 11.45$\pm$0.04 \\
RTX 5070 & 0.79$\pm$0.00 & 1.26$\pm$0.00 & 1.90$\pm$0.00 & 1.47$\pm$0.00 & 4.69$\pm$0.00 & 0.50$\pm$0.00 \\
M3* & 8.34$\pm$0.03 & 12.49$\pm$0.01 & 12.13$\pm$0.02 & 18.45$\pm$0.02 & 32.72$\pm$0.04 & 5.92$\pm$0.07 \\
\bottomrule
\end{tabular}
}

The mesh--ray intervals do not overlap on any device. The M3 interval is now honest: this collector kept the CPU sample arrays.

\subsection{Preparation cost}\label{sec:prep}\label{preparation-cost}

Per-frame numbers exclude one-time work. The 780M run gives the following CPU wall times in milliseconds (maximum over its twelve scenes):

\pandocbounded{\includegraphics[width=\columnwidth]{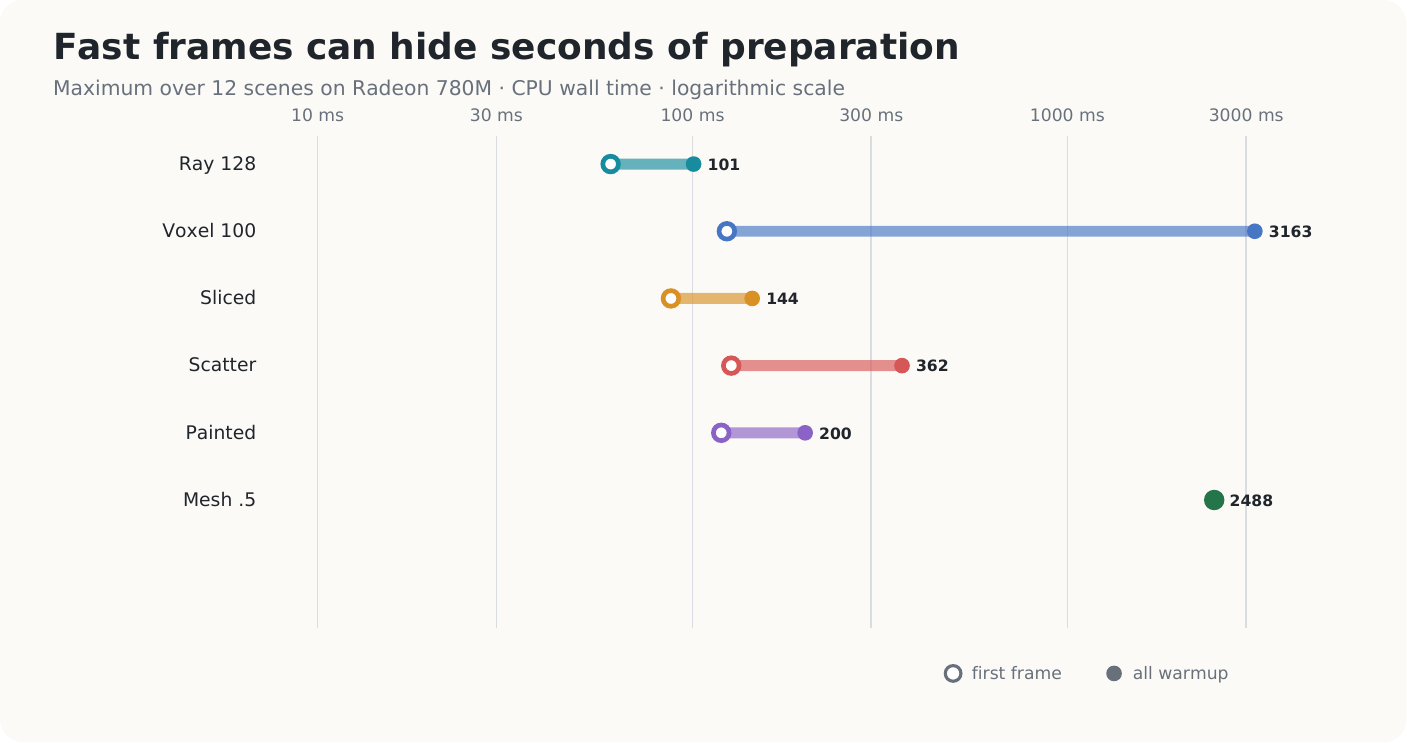}}

\begin{longtable}[]{@{}llll@{}}
\toprule\noalign{}
method & setup & first frame & warmup \\
\midrule\noalign{}
\endhead
\bottomrule\noalign{}
\endlastfoot
RayTraced & 9 & 60 & 101 \\
RayVoxel & 12 & 124 & \textbf{3163} \\
Sliced & 17 & 88 & 144 \\
Scattered & 10 & 127 & 362 \\
Painted & 18 & 119 & 200 \\
Mesh q=0.5 & 10 & \textbf{2464} & 2488 \\
\end{longtable}

\passthrough{\lstinline!setup!} builds pipelines and uploads the terrain texture; \passthrough{\lstinline!first frame!} adds whatever the method builds lazily; \passthrough{\lstinline!warmup!} covers every pre-timing frame. The two methods that pay anything substantial pay it differently. The mesh fits its triangulation once, on the CPU, in a single blocking 2.5 s at q=0.5 --- a load-time cost that a level cannot be entered without. The voxel grid bakes incrementally under a per-frame texel budget, spreading 3.2 s across frames that are individually playable but render through terrain the bake has not reached yet.

Neither is visible in a steady-state frame time, and for a level-loading budget the difference between them matters more than the per-frame gap.

\subsection{Editing and retained method data}\label{sec:edit}\label{editing-and-retained-method-data}

\pandocbounded{\includegraphics[width=\columnwidth]{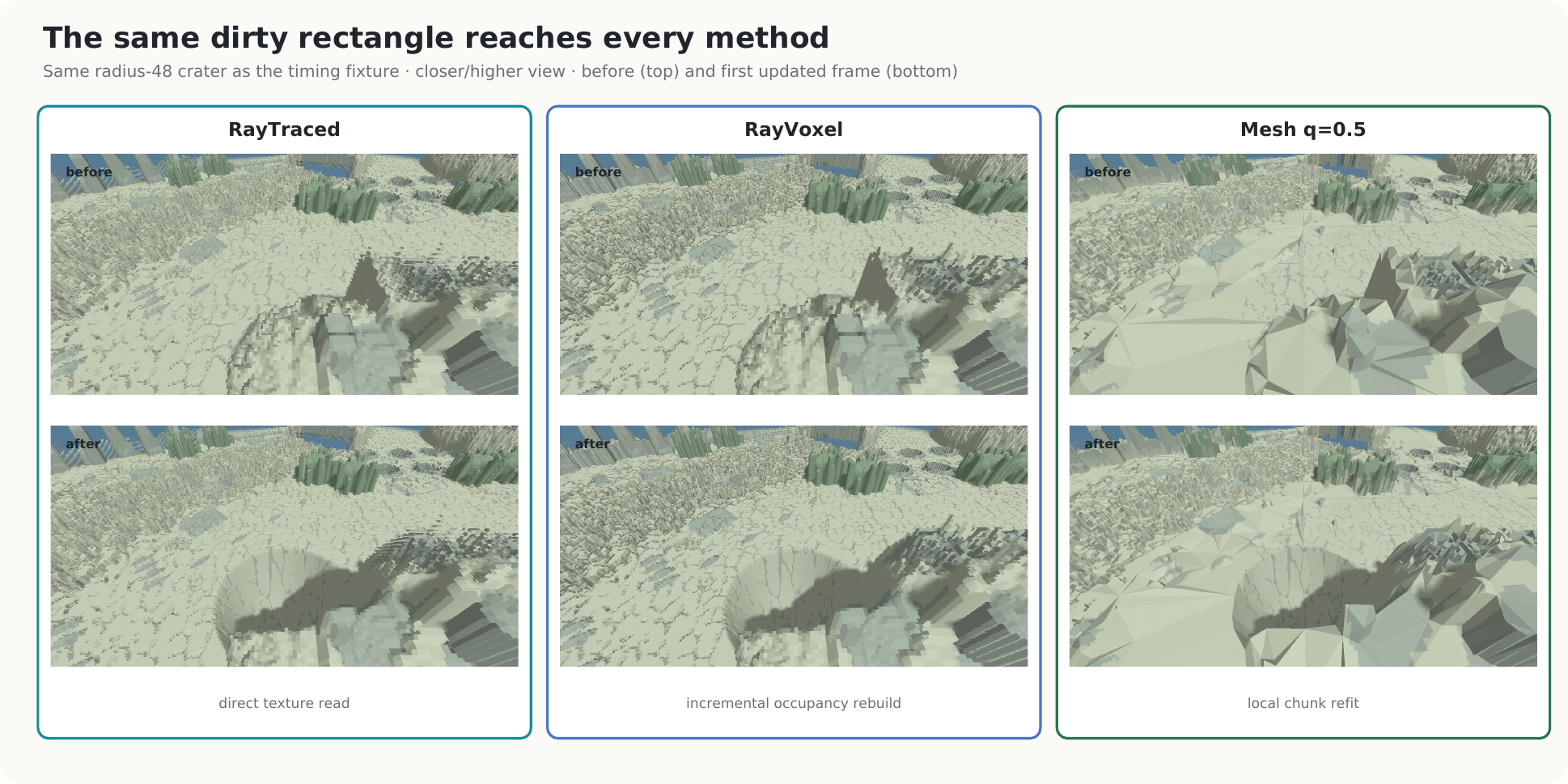}}

The headless \passthrough{\lstinline!--dig!} path removes altitude and upper-layer metadata inside a radius-48 crater and submits the same dirty rectangle the interactive game would. The camera stands south of the crater at pitch --30° (Section~\ref{sec:edit-protocol}). Figure 5.4 shows the unedited view and the first updated frame for the three update classes: RayTraced reads the new texture directly, RayVoxel rebuilds affected occupancy, and Mesh refits dirty chunks. All six methods receive the same edit; the figure shows one of each class.

Times below are medians on the Radeon 780M. \passthrough{\lstinline!first edit!} has five independent samples of the first post-edit frame. CPU is submit-and-wait (upload + any refine + GPU work + round trip): that is the latency a player waits. GPU is the timestamped draw of that same frame. \passthrough{\lstinline!steady!} is the median of ten frames from a fresh build of the already-edited level.

{\footnotesize\setlength{\tabcolsep}{3pt}
\begin{tabular}{@{}lrrrrrrl@{}}
\toprule
method & CPU$_0$ & CPU$_1$ & $\Delta$C & GPU$_0$ & GPU$_1$ & $\Delta$G & match \\
\midrule
Ray 128 & 6.51 & 8.70 & +2.19 & 5.59 & 7.45 & +1.86 & 1 frame \\
Voxel 100 & 7.54 & 9.00 & +1.46 & 6.60 & 6.97 & +0.37 & 1 frame \\
Sliced & 7.79 & 10.97 & +3.18 & 6.91 & 9.60 & +2.69 & 1 frame \\
Scattered & 8.41 & 10.16 & +1.75 & 7.37 & 8.65 & +1.28 & 1 frame \\
Painted & 17.89 & 22.72 & +4.83 & 16.87 & 21.04 & +4.17 & 1 frame \\
Mesh .5 & 6.07 & 15.94 & +9.88 & 4.77 & 6.07 & +1.30 & not by 16 \\
\bottomrule
\end{tabular}
}

On this integrated GPU the first updated frame also costs extra GPU time for the methods that upload or rebuild, because the dirty work is not hidden behind a fast discrete queue. The mesh is still the outlier: its CPU $\Delta$ is 9.88 ms while its extra GPU time is only 1.30 ms. Across the five devices Mesh CPU $\Delta$ is 5.0--9.9 ms (7900 XT 5.03, M3 5.98, RTX 5070 7.31, Intel 9.83, 780M 9.88). The five direct or regularly rebuilt methods match a fresh edited build on the first updated frame on every adapter. The mesh does not: at 16 frames it still differs by 0.0016\% hit/miss classification, 0.22 u p95 depth, and 0.78--0.80 levels of 8-bit color MAE, identically on all five devices. That remainder sits inside the q=0.5 tolerance; exact rebuild equivalence is nevertheless a stronger property than this implementation provides.

The publication batch is a clean checkout of \passthrough{\lstinline!5586e2a!} (\passthrough{\lstinline!terrain-paper!}) on all five machines.

At 1280$\times$800 the same run reports explicit, persistent, method-specific allocation. Zero means the method adds no persistent data beyond the shared renderer resources excluded in Section~\ref{sec:edit-protocol}.

\begin{longtable}[]{@{}lrr@{}}
\toprule\noalign{}
method & GPU data (MiB) & CPU data (MiB) \\
\midrule\noalign{}
\endhead
\bottomrule\noalign{}
\endlastfoot
RayTraced 128 & 0 & 0 \\
RayVoxel (4,8,2) & 18.29 & \textless0.01 \\
Sliced & 0 & 0 \\
Scattered & 3.91 & 0 \\
Painted & \textless0.01 & 0 \\
Mesh q=0.5 & 318.7 & 534.7 \\
\end{longtable}

RayVoxel\textquotesingle s tuned coarse grid is only 18.29 MiB; its 153 MiB production grid remains a distinct, disclosed configuration (Section~\ref{sec:tuning}). Scattered\textquotesingle s 3.91 MiB is one 32-bit value per output pixel and therefore scales with resolution. The mesh retains three LODs plus every chunk\textquotesingle s live triangulation. These figures are explicit payload sizes, not peak process or driver memory.

\subsection{Fit cost}\label{sec:fit}\label{fit-cost}

Fostral, triangles against a full grid mesh:

\begin{longtable}[]{@{}lllll@{}}
\toprule\noalign{}
quality & max error & vertices & triangles & reduction \\
\midrule\noalign{}
\endhead
\bottomrule\noalign{}
\endlastfoot
0.0 & 16 & 1.74 M & 3.39 M & 19.8$\times$ \\
0.25 & 8 & 2.31 M & 4.50 M & 14.9$\times$ \\
0.5 & 4 & 4.28 M & 8.47 M & 7.9$\times$ \\
0.75 & 2 & 7.14 M & 14.1 M & 4.7$\times$ \\
1.0 & 1 & 11.3 M & 22.4 M & 3.0$\times$ \\
\end{longtable}

Against roughly 80$\times$ for a smooth synthetic surface at comparable tolerance (our own control, not a literature figure), and 45--182$\times$ for the single-layer stock worlds (Section~\ref{sec:survey}). The stock worlds are the better control than an external elevation model would be: they vary only content, holding encoding, quantisation, texel scale and authoring pipeline fixed, so the comparison isolates one variable rather than four.

\subsection{Tuning}\label{sec:tuning}\label{tuning}

Each method was swept over its own quality knob and given the cheapest setting within one percentage point of its own best error, so no method is charged for a setting that buys nothing or credited with speed it reaches only by being wrong. Fostral, three viewpoints at the horizon, 400x260, view distance 600. The CPU reference was re-swept after correcting its screen-X basis and far-plane convention. The mesh received an additional 1280$\times$800 pass because its wall coverage was not resolved at tuning resolution. The complete sweeps are recorded in \passthrough{\lstinline!paper/tuning.md!}.

{\small\setlength{\tabcolsep}{4pt}
\begin{tabular}{@{}lllll@{}}
\toprule
method & knob & swept & chosen & error \\
\midrule
RayTraced & steps & 16--256 & \textbf{128} & 1.2\% \\
Painted & --- & --- & --- & 0.1\% \\
Sliced & slices & 32--512 & \textbf{512} & 6.5\% \\
Scattered & density & 1--4 & \textbf{4,4,4} & 39.2\% \\
RayVoxel & grid, steps & 2 grids $\times$ 40--400 & \textbf{4,8,2 / 100} & 0.3\% \\
Mesh & tolerance & q 0.0--1.0 & \textbf{q=0.5} & 0.3\% at 1280$\times$800 \\
\bottomrule
\end{tabular}
}

Four of the results are worth stating.

\textbf{The slicer knob changes the error\textquotesingle s kind.} From 32 to 512 slices, cost rises 0.31 → 1.54 ms while see-through falls 15.5\% → 3.0\%. Speckle is non-monotonic because additional slices convert missing spans into isolated wrong pixels before coverage becomes dense enough; the combined error falls 16.7\% → 6.5\%, so the rule selects 512.

\textbf{The voxel step budget is a property of the fixture, not the method.} 40 steps leaves 2.5\% combined error, while 100 reaches 0.3\%; 200 and 400 remain at 0.3\% and only add work. The rule therefore selects 100. A step budget tunes the longest sightline the viewpoints put in frame; change the viewpoints and it needs re-tuning, which is what the protocol\textquotesingle s tuning pass is for.

\textbf{Resolution changes whether the reference can resolve mesh quality.} At 400$\times$260 q=0.0 and q=0.25 both have 1.8\% combined error, so the rule initially chooses q=0.5 as the first point within one point of the 0.7\% best. At 1280$\times$800, q=0.0 leaves 3.9\%, q=0.25 leaves 0.8\%, and q=0.5--1.0 leave 0.3\%. The one-point rule would pick q=0.25. We publish \textbf{q=0.5} instead: it is the coarsest setting that matches the full-resolution error floor, and it lets the comparison, teaser, and video share one mesh. The hangar wall that q=0.0 drops is recovered at this setting.

\textbf{The ray marcher selects 128 steps.} Across the final horizon scenes, 16 → 256 steps costs 0.24 → 1.29 ms while total error falls 6.9\% → 0.6\%. The 128-step setting is the cheapest within one point of the best (1.2\%). A cheaper 64-step setting is not substituted for the selected one.

A fifth result is about configuration rather than the method: the voxel tracer\textquotesingle s production grid (2,4,1) needs 153 MB of storage buffer and did not fit the software rasterizer used for tuning. The selected comparison runs the (4,8,2) grid even though all five hardware devices can accommodate the production grid. The reported comparison therefore characterises the tuned coarse configuration, not the renderer\textquotesingle s shipping configuration; the memory requirement is part of the configuration disclosure rather than an unreported advantage.

\subsection{Fastest is not free}\label{sec:choice}\label{fastest-is-not-free}

At the selected quality point Mesh q=0.5 is the lowest mean on every device, and it is also in the coherent group of Section~\ref{sec:pitch}. Frame time is therefore not the remaining argument against it. The remaining costs are the ones the timing table hides: 2.5 s of blocking fit, 319 MiB of GPU buffers and 535 MiB of CPU triangulation, and a 5--10 ms history-dependent refit that never quite matches a fresh build.

RayTraced 128 is the other production-shaped choice. It is second in the timing table, needs no extra memory, and sees a crater on the next draw. Its fixed sampling budget is still visible as blocky close-up detail, and grazing rays remain the worst possible workload. The selected 128-step setting is the cheapest point within the tuning rule, not a quality match for interpolated triangles.

The choice is consequently workload-level. RayTraced is compelling when load time, memory, and exact first-frame edits dominate. Mesh is the stronger candidate when close-range quality, grazing views, or a triangle-mesh collider matter enough to amortise fitting and local refits. The new measurement is that those mesh advantages no longer have to be bought with a slower frame.

\section{Findings}\label{sec:findings}\label{findings}

\subsection{The remaining depth offset}\label{sec:offset}\label{the-remaining-depth-offset}

After the reference corrections in Section~\ref{sec:reference}, \passthrough{\lstinline!covers-sky!} at all three --90° scenes is exactly 0.0\% for every method, and the pitch means remain at or below 0.2\%. The selected coherent methods leave at most 0.9\% uncovered at the horizon.

One limitation stays visible in the numbers. In extremely wide off-axis top-down pixels, otherwise agreeing methods share median absolute offsets of roughly 8--14 world units against the CPU point marcher. Halving its step did not remove the offset, so this is a point-sampling or cell-boundary convention rather than insufficient convergence. Absolute depth is therefore a diagnostic, not evidence for sub-unit ranking. Quality claims rest on bidirectional coverage, coherence, visual agreement, and relative agreement between independent renderers.

\subsection{The multi-layer encoding, not the terrain, sets the fit cost}\label{sec:survey}\label{the-multi-layer-encoding-not-the-terrain-sets-the-fit-cost}

The obvious reading of a 14.9$\times$ reduction where a smooth surface gives \textasciitilde80$\times$ is that hand-authored terrain is simply harder to fit. The ten shipped worlds let us test that directly: same engine, same encoding, same 8-bit quantisation, same texel scale, same authoring tools, varying only content. Fitted at identical tolerance:

\pandocbounded{\includegraphics[width=\columnwidth]{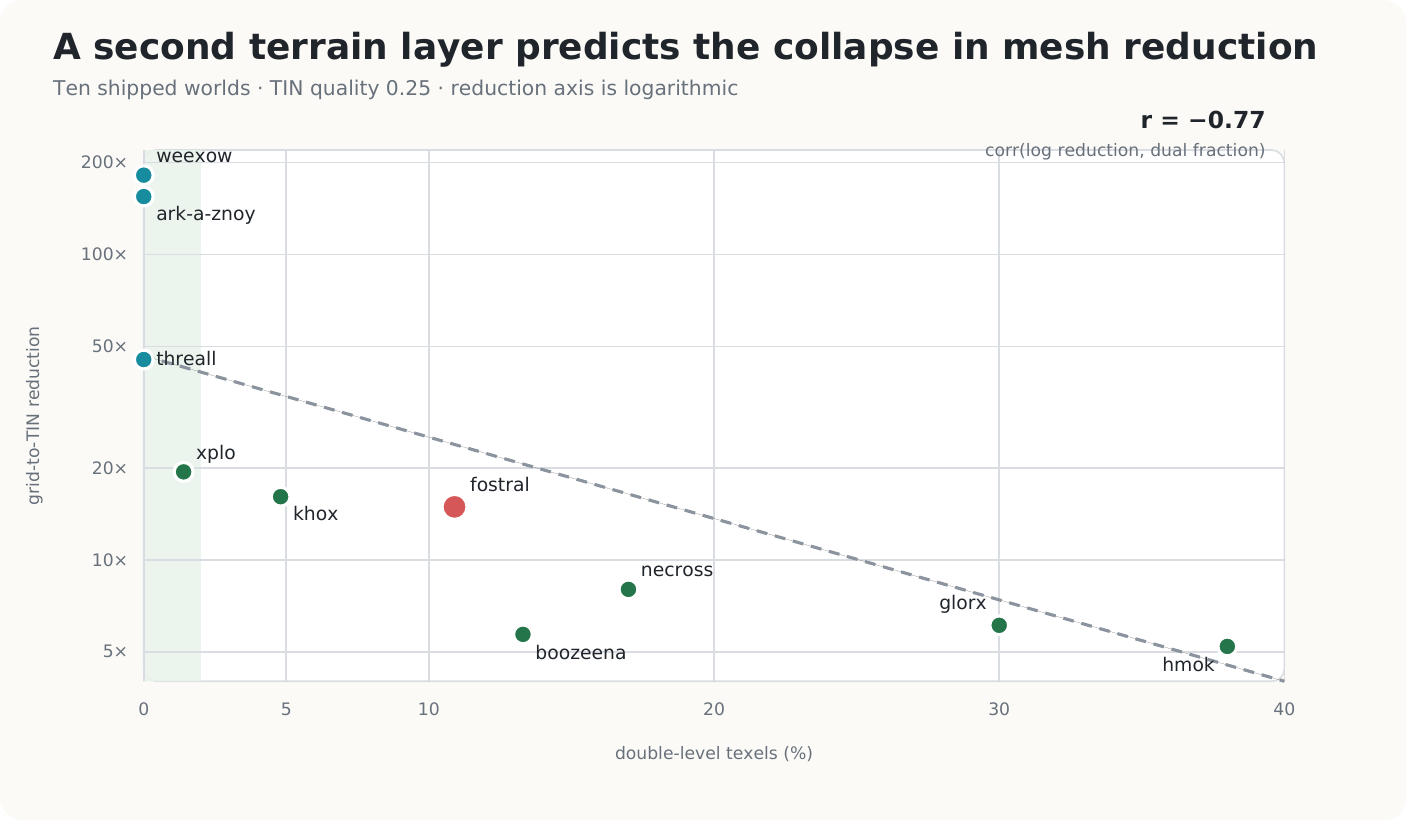}}

{\footnotesize\setlength{\tabcolsep}{3pt}
\begin{tabular}{@{}lrrrrrrr@{}}
\toprule
level & texels & tris & red. & slab & dual & $r_f$ & $r_s$ \\
\midrule
weexow & 4.2\,M & 0.05\,M & 182.0$\times$ & 0.0\% & 0.0\% & 1.84 & 1.84 \\
ark-a-znoy & 4.2\,M & 0.05\,M & 154.8$\times$ & 0.0\% & 0.0\% & 5.52 & 5.52 \\
threall & 4.2\,M & 0.19\,M & 45.3$\times$ & 0.0\% & 0.0\% & 2.85 & 2.85 \\
xplo & 8.4\,M & 0.87\,M & 19.4$\times$ & 17.0\% & 1.4\% & 7.54 & 8.80 \\
khox & 4.2\,M & 0.52\,M & 16.1$\times$ & 9.3\% & 4.8\% & 6.94 & 7.71 \\
fostral & 33.6\,M & 4.50\,M & 14.9$\times$ & 35.6\% & 10.9\% & 3.44 & 6.54 \\
necross & 33.6\,M & 8.41\,M & 8.0$\times$ & 42.5\% & 17.0\% & 4.41 & 14.43 \\
glorx & 33.6\,M & 10.97\,M & 6.1$\times$ & 37.2\% & 30.0\% & 6.53 & 14.66 \\
boozeena & 4.2\,M & 1.46\,M & 5.7$\times$ & 41.5\% & 13.3\% & 3.44 & 25.15 \\
hmok & 4.2\,M & 1.61\,M & 5.2$\times$ & 47.2\% & 38.0\% & 2.71 & 18.77 \\
\bottomrule
\end{tabular}
}

{\footnotesize $r_f$ is \texttt{rough(floor)}; $r_s$ is \texttt{rough(surface)}.}

\passthrough{\lstinline!rough(floor)!} is the mean absolute discrete Laplacian of the \passthrough{\lstinline!low!} layer alone --- the terrain\textquotesingle s own curvature, blind to the second layer. \passthrough{\lstinline!rough(surface)!} is the same measure on the composite surface the fitter sees.

The hypothesis fails. Across all ten worlds, correlation of log reduction against \passthrough{\lstinline!rough(floor)!} is \textbf{--0.17}; against the double-level texel fraction it is \textbf{--0.77}, and against \passthrough{\lstinline!rough(surface)!} --- the data the fitter actually sees, curvature the slab edges put there --- it is \textbf{--0.82}. (The share of the fitted triangles that end up on slab surfaces correlates at --0.89, which is the fit\textquotesingle s own account of where its budget went rather than an independent predictor.) Relief does not predict the fit cost; the second layer does. \passthrough{\lstinline!ark-a-znoy!} makes the floor-relief hypothesis fail in the other direction too: it is \emph{three} times as rough as \passthrough{\lstinline!weexow!} on the floor (5.52 against 1.84) and compresses essentially as well (154.8$\times$ against 182.0$\times$), because neither world has a slab.

The clearest case is \passthrough{\lstinline!hmok!}. Its floor is \emph{smoother} than \passthrough{\lstinline!threall!}\textquotesingle s (2.71 against 2.85), and \passthrough{\lstinline!threall!} compresses 45.3$\times$. \passthrough{\lstinline!hmok!} manages 5.2$\times$ --- nine times worse on flatter ground --- because 47\% of its triangles are slab. Its composite roughness is seven times its floor roughness, and all of that excess is the encoding.

So the honest claim is not that authored terrain defeats greedy TIN. Single-layer authored terrain compresses 45--182$\times$ in our own controls, which confirms the fitter itself is not the weak link without importing a ratio from a differently sampled elevation model. What defeats it is a second layer whose altitudes are structural rather than continuous. Section~\ref{sec:boundary} is the mechanism.

\subsection{A quarter of the vertex budget goes to one discontinuity}\label{sec:boundary}\label{a-quarter-of-the-vertex-budget-goes-to-one-discontinuity}

Counting what drove each insertion, Fostral at quality 0.25:

\begin{longtable}[]{@{}ll@{}}
\toprule\noalign{}
driver & share \\
\midrule\noalign{}
\endhead
\bottomrule\noalign{}
\endlastfoot
\passthrough{\lstinline!low!} (the floor) & 53.5\% \\
slab interior & 18.8\% \\
\textbf{single/double-level boundary} & \textbf{23.3\%} \\
chunk-border simplification & 4.4\% \\
\end{longtable}

A single-level texel reports \passthrough{\lstinline!mid = high = low!}, so both step by the full slab thickness across a region edge, and the error metric chases a discontinuity no tolerance can satisfy --- it only shrinks triangles toward texel size along every boundary. The absolute cost is near-constant in quality (338 k insertions at q=0, 396 k at q=1) while the floor\textquotesingle s grows 444 k → 6.1 M. That signature --- flat in tolerance --- distinguishes a geometric feature from a fit converging, and generalises to any error-driven fit over a field with embedded discontinuities.

Constraining the triangulation to the region outline would remove it, and is the same change that would stop straddling triangles dropping the slab.

\subsection{Coverage metrics cannot see over-drawing}\label{sec:overdraw}\label{coverage-metrics-cannot-see-over-drawing}

Coverage alone cannot fail a renderer that draws too much. A mesh that interpolates a slab across region boundaries covers every pixel that should be covered, and a see-through score treats that invented roof as correct. That is why Section~\ref{sec:metrics} splits coverage into see-through and covers-sky and adds a depth-coherence score: the hangar wall dropped at q=0.0 is recovered by comparing depth and by looking at the image, not by counting covered pixels.

\section{Limitations}\label{sec:limits}\label{limitations}

\begin{itemize}
\tightlist
\item
  One engine and one terrain format. The ten stock worlds span a factor of 35 in fit cost and isolate the multi-layer variable cleanly, but every one of them is Vangers data. Whether the mechanism in Section~\ref{sec:boundary} generalises to other discontinuous auxiliary fields is argued, not measured.
\item
  All rendering comparisons use Fostral, published under CC BY-SA 4.0 by Association K-D Lab. The nine additional survey worlds are not in that grant; those rows require a lawfully obtained game copy.
\item
  The hardware batch covers four Vulkan devices and one Metal device, but not D3D12, WebGL2, mobile-class adapters, or multiple driver versions per adapter. It establishes native WebGPU execution and image agreement across those backends, not a survey of every WebGPU implementation. The browser smoke test is a path-to-the-web check of one route, not a six-method web comparison.
\item
  Frame timing is per-frame latency, not pipelined throughput: each frame is submitted and awaited in isolation, so nothing overlaps. The Vulkan timestamps make those numbers GPU work rather than round trip, but they do not make them a frame rate. Metal uses CPU submit-and-wait because its encoder timestamps failed the bracketing sanity check; those values are useful within the M3 rows but cannot be compared directly with Vulkan.
\item
  Tuning uses one fixed three-scene horizon fixture. It selects 128 height-field steps, 100 voxel steps, and q=0.5 at publication resolution; changing the view-distance or scene distribution can select another operating point (Section~\ref{sec:tuning}).
\item
  The edit experiment covers one crater shape and one location on all five publication adapters. It establishes first-frame visibility, CPU and GPU update cost, and mesh history dependence. A different crater or a gameplay-shaped destruction pattern is not measured.
\item
  Explicit method payloads are accounted for, but portable wgpu does not expose opaque driver allocations or a backend-independent peak heap. Chunk streaming is not implemented, so "runs on low-end devices" describes the pipeline, not its memory budget.
\end{itemize}

\section{Conclusion}\label{sec:conclusion}\label{conclusion}

Six terrain renderers that look interchangeable from the original game\textquotesingle s top-down camera behave differently once the same authored data is viewed at eye level. Horizontal slices expose bands and point scattering exposes incoherent pixels. At the selected quality point the mesh has the lowest mean frame time on every measured adapter, and it stays in the coherent group at the horizon. Close detail remains blocky on the marchers. The mesh\textquotesingle s remaining cost is memory, a 2.5 s fit, and a history-dependent refit --- not the frame.

The larger result is about the data and the measurement. Single-layer worlds fit by 45--182$\times$, while the structural second layer, not floor relief, predicts the collapse in reduction; nearly a quarter of Fostral\textquotesingle s vertex insertions serve one layer-boundary discontinuity. The five direct or regularly rebuilt methods reproduce a fresh edited build on their first updated frame, but the selected insertion-only mesh retains a small history-dependent difference. Making that mesh editable also exposes its retained cost: 319 MiB of explicit GPU geometry and 535 MiB of CPU triangulation. A credible comparison needs bidirectional coverage, coherence, inter-method agreement, equal tuning, explicit preparation costs, and visual parity checks in addition to a timing table. It also needs to treat post-edit maintenance as a first-class result: a static hierarchy can win a frame and still fail the workload if terrain destruction forces a reload. Conversely, a method can win the timing table and still lose the application if its quality budget is view-dependent or its representation cannot be reused by the rest of the engine.

\subsection*{Acknowledgements}

I thank Association K-D Lab for \textit{Vangers} and for publishing the Fostral
world data that this comparison uses; Yury Zhuravlev for maintaining the
open-source Vangers tree; and the players and other maintainers who have
kept the game alive for nearly three decades.

Large language models --- OpenAI Codex, Anthropic Claude, and xAI Grok ---
assisted with drafting, editing, literature search, and work on the
evaluation harness. I reviewed every claim, number, and citation; the
remaining errors are mine.

\small
\bibliographystyle{jcgt}
\bibliography{references}

\section*{Index of Supplemental Materials}
Supplemental video: \texttt{anc/terrain-methods.mp4}
(\texttt{tools/render-paper-video.py}; portal $(1176,11567)$,
eye height 180, yaw $308^\circ$, 520 units).
Engine and harness: \href{https://github.com/kvark/vange-rs}{https://github.com/kvark/vange-rs},
tag \texttt{terrain-paper}.
Fostral is not redistributed; the harness fetches Association K-D Lab's
CC~BY-SA~4.0 tree at commit \texttt{f1ad7d7}.

\section*{Author Contact Information}

Dzmitry Malyshau\\
Independent Researcher\\
\href{mailto:kvark@fastmail.com}{kvark@fastmail.com}\\
ORCID \href{https://orcid.org/0009-0005-6410-4276}{0009-0005-6410-4276}

\vspace{2mm}\hrule\vspace{2mm}
{\small
\noindent Malyshau, Six Ways to Draw Vangers with WebGPU: Real-Time Rendering of Editable Multi-Layer Height Fields, submitted to \textit{Journal of Computer Graphics Techniques (JCGT)}.

\vspace{1.5mm}
\noindent \copyright~2026~Malyshau (the Authors).
The Authors provide this document (the Work) under the Creative Commons CC~BY-ND~4.0 license available online at \href{http://creativecommons.org/licenses/by-nd/4.0/}{http://creativecommons.org/licenses/by-nd/4.0/}.
The Authors further grant permission for reuse of images and text from the first page of the Work, provided that the reuse is for the purpose of promoting and/or summarizing the Work in scholarly venues and that any reuse is accompanied by a scientific citation to the Work.

\vspace{1mm}
\noindent\includegraphics[width=2.0cm]{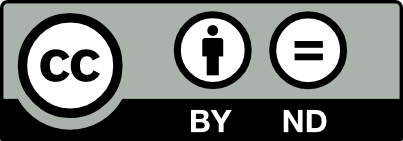}
}

\end{document}